\documentclass[preprint,showpacs,preprintnumbers,amsmath,amssymb]{revtex4}

\usepackage{graphicx}
\usepackage{dcolumn}
\usepackage{bm}

\begin{document}

\vspace{5mm}

\newcommand{\goo}{\,\raisebox{-.5ex}{$\stackrel{>}{\scriptstyle\sim}$}\,}
\newcommand{\loo}{\,\raisebox{-.5ex}{$\stackrel{<}{\scriptstyle\sim}$}\,}

\title{Formation of exotic baryon clusters in ultra-relativistic heavy-ion 
collisions.}

\author{A.S.~Botvina$^{1,2}$, J.~Steinheimer$^{1}$, M.~Bleicher$^{1,3}$}

\affiliation{$^1$Frankfurt Institute for Advanced Studies and ITP J.W. Goethe 
University, D-60438 Frankfurt am Main, Germany} 
\affiliation{$^2$Institute for Nuclear 
Research, Russian Academy of Sciences, 117312 Moscow, Russia} 
\affiliation{$^3$John von Neumann Institut f\"ur Computing (NIC), 
J\"ulich Supercomputing Centre, FZ J\"ulich, D-52425 J\"ulich, Germany}

\date{\today}

\begin{abstract}

Recent experiments at RHIC and LHC have demonstrated that there are excellent 
opportunities to produce light baryonic clusters of exotic matter 
(strange and anti-matter) in ultra-relativistic ion collisions. Within 
the hybrid-transport model UrQMD we show that the coalescence mechanism can 
naturally explain the production of these clusters in the ALICE experiment 
at LHC. As a consequence of this mechanism we predict the rapidity domains 
where the yields of such clusters are much larger than the observed one at 
midrapidity. This new phenomenon can lead to unique methods for producing 
exotic nuclei. 

\end{abstract}

\pacs{25.75.-q , 25.75.Dw , 25.75.Ld , 21.80.+a }

\maketitle

\section{Introduction}

In relativistic nuclear collisions an abundance of new particles 
consisting of all kind of quark and anti-quark flavors is produced. 
During the late stage of the collision these particles can interact in 
secondary processes and produce novel clusters containing several baryons. 
In this case, promising studies of fragmentation reactions probing 
the limits in isospin space of light nuclei, exotic nuclear states, 
anti-nuclei, and multiple strange nuclei are feasible. 
Recently very encouraging results on the formation of exotic clusters 
come from experiments at relativistic colliders: For example, STAR at 
RHIC \cite{star,ygma-nufra}, and ALICE at LHC 
\cite{alice-3LH,camerini-nufra} have observed 
hyper-tritons and anti-hyper-tritons. Experimental programs to search for 
more heavy exotic nuclear species are now underway \cite{YGMa13,Don15}. 
Therefore, a theoretical understanding of these phenomena is necessary. 
Transport models have been used to successfully describe many 
observables, including strangeness production at 
intermediate energies \cite{Bra04,Har12,Bas98,Ble99,Bot17}. At very high 
energy most of the 
state-of-art hybrid models apply a hydro-dynamical expansion of the hot 
and dense matter and a subsequent microscopic transport approach 
to describe the hadronic rescattering 
(see, e.g., for UrQMD, 
Ref.~\cite{Steinheimer:2015msa,Steinheimer:2017vju}). 
In the framework of microscopic 
transport models a coalescence prescription for the formation of the 
composite clusters can be naturally applied \cite{Ton83,Gyu83,Nag96,Bot15}. 
In this 
paper we demonstrate the effectiveness of the transport-plus-coalescence 
approach for the description of data at LHC energies. 
Important predictions for the 
future research of baryon clusters in the ultra-relativistic 
heavy-ion collisions are also presented.

\section{Models for production of light clusters at relativistic collisions}

Thermal and coalescent mechanisms to produce complex nuclei in high 
energy collisions have been discussed in previous works (see, e.g., 
\cite{And11,Ste12}). The thermal models allow for a good description of 
the particle production yields, for example, in the most central 
collisions \cite{And13,Sta14}. For this reason we believe 
that the produced particles do widely populate the available reaction 
phase space, and this should be taken into account in any interpretation 
of the data. Only the lightest clusters, with mass numbers A$\loo$3--4, 
can be noticeably produced in this case because of the very high 
temperature of the fireball (T$\approx$160 MeV). However, the pure 
thermal models can not describe the energy spectra of particles 
and their flows. Also in non-central collisions the dynamics and secondary 
interactions in the projectile and target residues will influence the 
nucleon clusters (fragments) production. As was shown, the thermal 
and coalescence descriptions are naturally connected: In particular, 
there is a relation between the coalescent parameter, density, temperature, 
and binding energies of the produced clusters \cite{Neu03}. 
In the following we consider the dynamical transport and coalescence 
mechanisms, because they have predictive power for many observables. 
There were also numerous discussions that even in central 
collisions of very high energy the coalescence mechanism, which 
assembles light fragments from the produced hyperons and
nucleons (including anti-baryons), may be essential 
\cite{star,ygma-nufra}.

The first reaction step should be the dynamical production of baryons 
which later on 
can be accumulated into clusters. The transport model Ultrarelativistic 
Quantum Molecular Dynamics (UrQMD) is quite successful in the description of 
a large body of data. In the standard formulation \cite{Bas98,Ble99} the 
model involves string formation and its fragmentation according to the 
PYTHIA model 
for individual hard hadron collisions. The current versions of UrQMD 
include up to 70 baryonic species (including their anti-particles), as 
well as up to 40 different mesonic species, which participate in binary 
interactions. This work is focused on very high energies and we employ the 
UrQMD transport model \cite{Petersen:2008dd} in the hybrid mode for the 
description of the dynamical evolution in central collisions. In this mode 
the propagation is composed of an ideal 3+1d fluid dynamical description for 
the 
dense phase, which is mainly compromised of a strongly interacting quark 
gluon plasma (QGP). The event-by-event initial state for the fluid 
dynamical evolution is calculated using the PYTHIA version implemented in 
the UrQMD model, where the starting time of the fluid dynamical evolution 
is set to $\tau_0=0.5$ fm/c. The equation of state, which governs the 
dynamical evolution has been discussed in 
detail in Ref.~\cite{Steinheimer:2011ea} and describes the transition from 
a hadronic system to the QGP as a smooth crossover at low baryon densities. 
Once the system dilutes, and the fluid dynamical 
description is no longer valid, the propagated fields are transformed 
into particles via a sampling of the Cooper-Frye equation 
\cite{Cooper:1974mv}. Here we explicitly conserve the net-baryon number, 
net-electric-charge, and net-strangeness as well as the total energy and 
momentum. After this transition all hadrons 
continue their evolution and may interact via the hadronic cascade part 
of the UrQMD model. This dynamical decoupling takes on the order of 
10--20 fm/c and has a significant influence on the observed hadron 
multiplicities \cite{Becattini:2012xb,Becattini:2016xct} and spectra 
\cite{Steinheimer:2015msa}, which is strongest for most central collisions. 
Consequently it has been shown that 
this model reasonable describes hadron spectra observed by the ALICE 
collaboration \cite{Steinheimer:2015msa}, in particular the proton 
spectra which are essential for the study of nuclei production.

The advantage of the Monte-Carlo transport final state description is that 
it provides 
event-by-event simulations of the baryon production. This is important for 
investigating correlation phenomena. The coalescent procedure is ideal for the 
description of the baryon accumulation into clusters on event-by-event basis. 
It was shown before that 
the coalescence criterion, which uses the proximity of baryons in momentum 
and coordinate space, is very effective in description of light nucleon 
fragments at intermediate energies \cite{Ton83,Gyu83,Nag96,Ste12}. 
After the dynamical stage described by UrQMD model we apply a generalized 
version of the coalescence model \cite{Bot15} for the coalescence of baryons 
(CB). In such a way it is possible to form primary 
fragments of all sizes, from the lightest nuclei to the heavy residues, 
including hypernuclei and other exotics within the same mechanism. 
It was previously found \cite{Bot15} that the optimal time for applying the 
coalescence (as a final state interaction) is around 40--50 fm/c after 
the initial collisions of heavy-ions, when the rate of individual 
inelastic 
hadron interactions decreases very rapidly. A variation of the time within 
this interval leads to an uncertainty in the yield around 10$\%$ for a fixed 
coalescence parameter. This is essentially smaller than the uncertainty in 
the coalescence parameter itself. The most important CB parameter is the 
maximum variance between velocities of baryons $v_{c}$ in a coalescent 
cluster. $v_{c}$ should be around $v_{c}\approx 0.1$c for the 
lightest clusters, to be consistent with their binding energy. This value is 
also supported by a comparison to experimental data at energies around 
1--10 A GeV \cite{Ton83}. 
We should note that our formulation of the coalescence model is microscopic, 
therefore, it takes into account all correlations and fluctuations of 
the particle production during the dynamical stage. For this reason we need 
a smaller coalescence parameter in order to describe the data than the 
parameters obtained in the analytical formulation of the coalescence 
\cite{gut76}. In principle, the coalescence to clusters with $A>4$ is also 
possible, however, these heavy clusters are 
expected to be excited and their following decay can be described with the 
statistical models \cite{Bot07,Bot13}. Usually such big primary 
fragments can be produced only in peripheral collisions from nuclear residues 
in the projectile and target rapidity region \cite{Bot15}. The advantage of 
the sequential approach (dynamics + coalescence + statistical decay) is the 
possibility to predict the correlations and fluctuations of the yields of 
all nuclei, including their sizes, with the rapidity, and with other 
produced particles. However, in the midrapidity region, because of a very 
large energy deposition, we expect the formation of small clusters only. 
In the following we concentrate on the LHC heavy-ion reactions, 
and on the latest results on light cluster production obtained by the ALICE 
collaboration \cite{alice13,alice16,alice-flow}. 


\section{Comparison with experimental data}

\begin{figure}[tbh]
\includegraphics[width=0.6\textwidth]{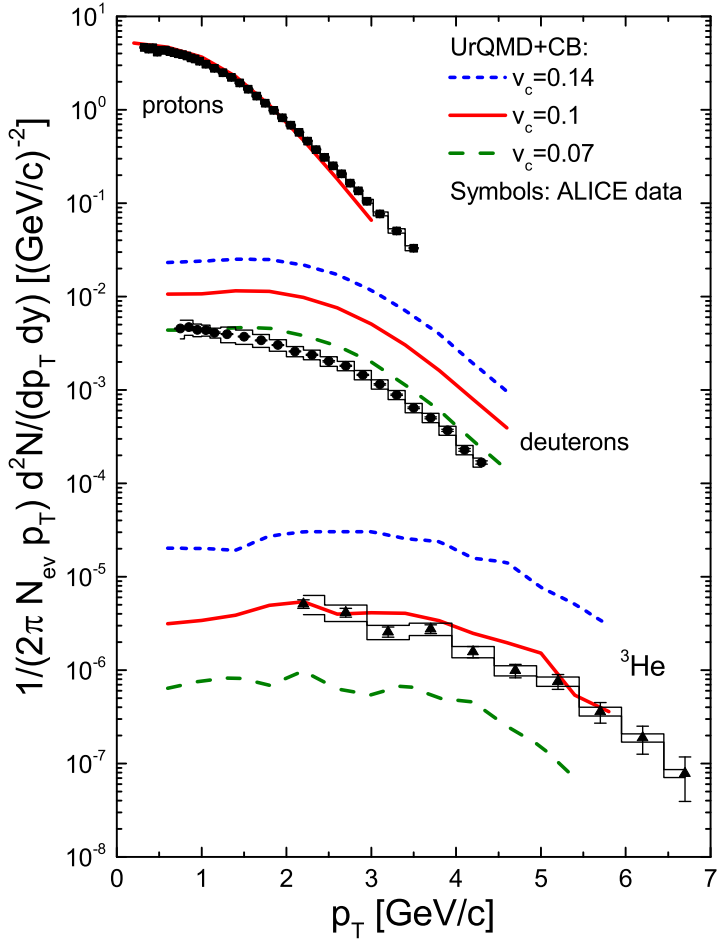}
\caption{\small{ (Color online) 
Transverse momentum spectra of protons, deuterons, and $^{3}$He 
particles in $^{208}$Pb on $^{208}$Pb collisions at 
the center of mass energy $\sqrt s$=2.76 TeV per nucleon. The 
symbols with systematical (thin boxes) and statistical errors are 
ALICE experimental data \cite{alice13,alice16}  
in the center-of-mass rapidity range from -0.5 to +0.5, and normalized 
per number of events for the 20\% most central events. The UrQMD coupled 
with coalescence of baryons (CB) calculations of the same particle spectra 
at the same conditions are shown by short-dashed (blue), solid (red) and 
long-dashed (green) lines 
for the corresponding coalescence parameters $v_c$ (see in the figure). 
}}
\label{fig1}
\end{figure}

We start with an analysis of the particle spectra as observed in the 
experiments. 
In Fig.~1 we show experimental data on transverse momentum distributions of 
protons, deuterons, and $^{3}$He particles measured at LHC by the ALICE group 
\cite{alice13,alice16}. The collisions of $^{208}$Pb on $^{208}$Pb have 
been performed at a center of mass energy of $\sqrt s$=2.76 TeV per nucleon. 
The yields in Fig.~1 are obtained for the central 
events (top 20\% of the maximum particle 
multiplicity) are normalized to the number of events. The rapidity range 
for detected particles was $y$=(-0.5 to +0.5) in the center of mass system. 
The experimental data are given by symbols inside boxes presenting 
the systematical uncertainties which are usually larger than the 
statistical ones. The statistical error bars are given if they 
are larger than the symbol sizes. This data presentation provides consistent 
information on yields and distributions of produced particles needed for 
verification of our models. 
The UrQMD hybrid calculations (including the hydro-dynamical evolution of 
matter) with the following CB calculations are shown by the lines. The 
different line styles depict variation of the coalescence parameter $v_c$ 
by 40\%. It is important that it is 
possible to reproduce very good the spectra of protons with UrQMD, since 
in the coalescence approach the yields of all clusters depend crucially 
on the baryon distributions. We should note that the yields at very high 
transverse momenta $P_T > 3-4$ A GeV are possibly dominated by jets, 
which are not currently included in the hydrodynamical evolution of the 
system. Therefore, 
we limit the fragments under study to $P_T \loo 2-3$ GeV per nucleon. 

One can see that the spectra of deuterons ($^2$H) and helium-3 ($^3$He) 
can be reasonably 
described with the coalescent parameters $v_c$=0.07c and $v_c$=0.1c, 
respectively. The larger value of $v_c$ for $^3$He is consistent with 
the larger binding energy of $^3$He in comparison with $^2$H. We note that 
the effect of the $v_c$ parameter is essentially bigger for large clusters. 
We could get a better agreement by tuning the coalescent parameters, 
however, this kind of phenomenological fitting is out of our theoretical 
study. It is more important that 
the form of the distributions is independent on $v_c$ in the wide range 
and corresponds to experimental distributions. 
This gives us a confidence to claim that the coalescence can naturally 
describe the production of these clusters.

\begin{figure}[tbh]
\includegraphics[width=0.6\textwidth]{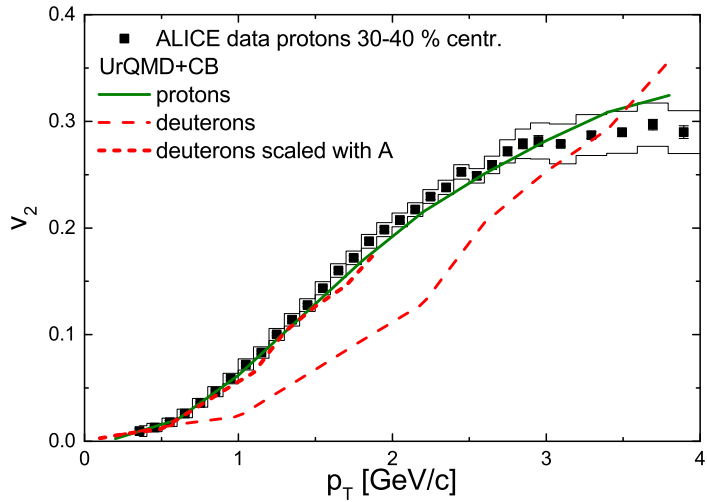}
\caption{\small{ (Color online) Elliptic flows ($v_2$) 
of produced protons and deuterons versus their 
transverse momenta. The reactions and the rapidity range are as in 
Fig.~1, however, the semicentral collisions (with the centrality range 
of 30\% -- 40\%) are selected. The ALICE experimental data \cite{alice-flow} 
for protons are the square symbols with errors within the thin boxes. 
The solid (green) and long-dashed (red) lines are the UrQMD and CB 
calculations for protons and deuterons respectively. 
The short-dashed red line presents the scaled distribution for 
deuterons (see the text). 
}}
\label{fig2}
\end{figure}

Another verification of the coalescence mechanism should come from 
angular distributions of the produced particles and their correlations 
respective to the reaction plane. 
We note that the angular (azimuthal) distribution of produced particles 
in the plane perpendicular to the beam axis is anisotropic with the 
corresponding maximum in the reaction plane. That is an expected consequence 
of the dynamical emission in such high energy collisions. 
A very informative observable is the 
elliptic flow $v_2$. Sometimes it is difficult to extract the reaction 
plane in the experiment, because of particle fluctuations in the 
collision events. In this case, particle correlation 
methods are used \cite{alice-flow}. For the present calculations we employ 
the reaction plane method in each collision, and, therefore, we 
can find $v_2$ for all particles by averaging their 
momenta perpendicular to the beam axis: 
$v_2 = \langle (P_x^2-P_y^2)/P_T^2 \rangle$, 
where $P_x$ is the momentum in the reaction plane, and 
$P_T^2=P_x^2+P_y^2$ is the transverse momentum. The averaging is done 
over all events containing these particles. 
It was shown that the reaction plane method provides results compatible 
with high-order event plane (correlation) methods \cite{Zhu05}. Therefore, 
$v_2$ trends versus $P_T$ should be a solid observable for comparison. 

We present $v_2$ measured by ALICE for protons in $^{208}$Pb on $^{208}$Pb 
reactions at $\sqrt s$=2.76 A TeV for semi-central collisions 
\cite{alice-flow} in Fig.~2. The semi-central events, which cover a 
centrality domain from 30\% to 40\% of the total particle 
multiplicity distribution, were used in this analysis. 
The UrQMD~+~CB calculations were performed under the same conditions for 
both protons and deuterons. One can see that the calculations 
describe the data for protons good, and they predict a rather different 
behaviour of $v_2$ versus $P_T$ for deuterons. However, our calculations 
lead to an interesting result: Namely, if we plot $v_2/A$ versus $P_T/A$ 
for protons (A=1) and deuterons (A=2), they are overlapping each other. 
Such a 'scaled' curve for deuterons is demonstrated by the short-dashed 
line in Fig.~2. This kind of 'scaling' of $v_2$ in the coalescence mechanism 
can be easy explained by the averaging procedure over the produced particles: 
When individual nucleons have nearly the same momenta the expression 
$(P_x^2-P_y^2)/P_T^2$ does not change after their clusterizing. However, 
the number of nucleons is by A  times larger than the number of 
clusters. The observation of such a coalescence 
scaling in experiments could be an additional verification of a pure 
coalescence mechanism. It is interesting that the scaling behavior has 
been observed in the experiments \cite{alice-flow}, however, in the 
elliptic flow of hadrons by taking into account the number of the 
constituent quarks. The quark coalescence was also discussed 
theoretically \cite{Mol03}. 
We believe this effect should be much stronger in our case of light 
nuclei, since the nuclear binding energy is much smaller than the 
nucleon masses and the scaling itself depends on the coalescence parameter 
rather weak.

\section{Predictions for the cluster rapidities}

\begin{figure}[tbh]
\includegraphics[width=0.6\textwidth]{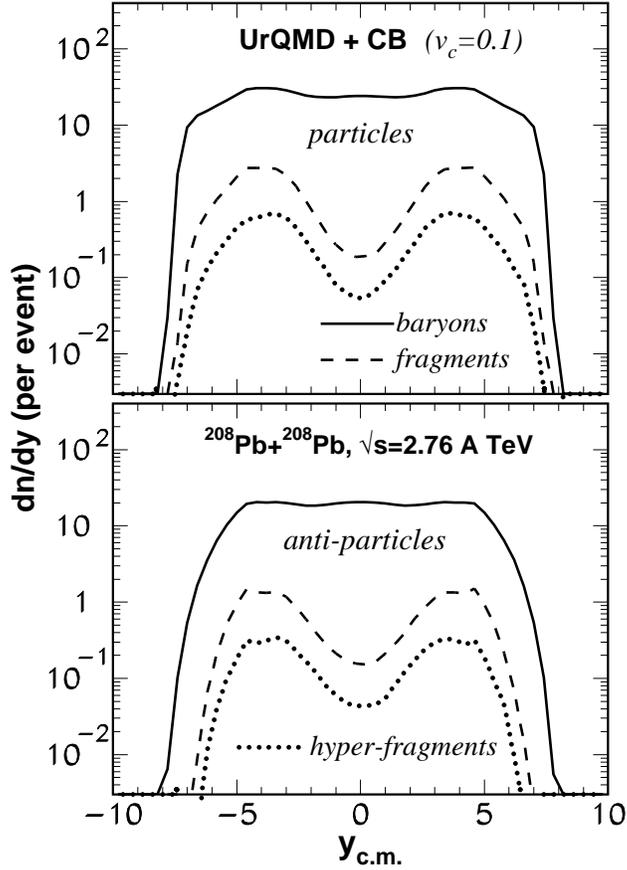}
\caption{\small{ The total rapidity distributions 
of produced particles normalized 
per one event. Top panel: normal baryons (solid line), all composite 
nucleon fragments (dashed line), and all composite hyper-fragments 
(thick dotted line).  Bottom panel presents the same but for 
anti-particles. The UrQMD and CB calculations are performed for 
$^{208}$Pb on $^{208}$Pb collisions at $\sqrt s$=2.76 A TeV overall impact 
parameters (minimal bias). 
}}
\label{fig3}
\end{figure}

Due to technical realization of the UrQMD hybrid model 
\cite{Steinheimer:2015msa} we have presented coalescence results from this 
model in the midrapidity range of central and semi-central Pb+Pb 
collisions only. It is however instructive to study also the rapidity 
dependence of cluster production, including peripheral collisions, 
as the produced systems properties may change essentially with rapidity. 
For example, special properties of nuclear matter near the hadron 
fragmentation region were discussed long ago \cite{Ani80}. It was suggested 
that the hadron matter of this region would have a significant 
non-zero net baryon number and high density \cite{Li:2017tcd}. 
Also the transverse momentum distributions of produced particles and 
fragmented nucleons may be different, that can have a significant effect 
on nuclei formation. In order to make an estimate on the rapidity 
distribution of nuclear clusters at the LHC we use the standard (cascade) 
version of UrQMD to generate baryons and their momenta. Then these are again 
used to form nuclei and hypernuclei via the CB model as described above. 
One should note here that the proton distributions fall more steeper with 
the transverse momentum than it was obtained in the hybrid version with 
hydrodynamics. Nevertheless, the main mechanisms of the particle production 
related to the secondary interactions remain the same, and the total 
particle and anti-particle yields are close. For this reason, the trends 
characterizing the modification of baryon momentum distributions with 
rapidity will be similar in the both versions. 

We demonstrate now predictions of the coalescent approach which are 
important for further investigations of the nuclear cluster formation 
in ultra-relativistic nuclear collisions. We have performed 
the UrQMD~+~CB calculations for all impact parameters 
with the minimum bias prescription. Fig.~3 shows the 
full rapidity distributions of baryons and obtained from them composite 
fragments of all sizes. For comparison, the top panel is for normal 
particles, and the bottom one is for anti-particles. Also we give 
separately fragments from nucleons and hyper-fragments which includes 
hyperons. 
We should note that in this and next figures we do not show the 
spectator nucleons and normal clusters composed from nucleons with 
rapidities around the projectile/target one (i.e., with 
$|y| \approx 8$). Slow participant nucleons may exist in this region 
and form clusters within the coalescence model. However, the full 
consideration requires 
a detail description of the excitation and de-excitation (via 
particle emission) of spectator residues, that is beyond the present 
paper. Moreover, these clusters can hardly be measured in present 
experiments because of very high rapidities.

For clarity, we have demonstrated results for one coalescence parameter 
$v_c = 0.1$, which is reasonable for the description of the data (Fig.~1).  
One can see a very broad distribution of the produced baryons in the 
rapidity. 
At such a high energy nearly the same amount of normal and anti-baryons 
are present at central rapidities. 
The broad rapidity distribution of the yields 
have already been discussed at 
intermediate collision energies \cite{Bot15,Ant16}. 
It is seen from Fig.~3 
that the production maximum for all composite fragments is shifted from 
midrapidity to the forward and backward region. In our case the wide 
maxima are located at the 
center-of-mass rapidities around +4 or -4. The reason of this 
phenomenon is in many secondary interactions and the energy loss 
during the hadron diffusion from midrapidity. 
An essential part of these interactions takes place 
between the newly produced species and the nucleons of projectile and target 
which did not interact in early times of the reaction. 
For this reason, both the energies and relative momenta of produced 
new baryons become smaller, therefore, it is easier for them to 
coalesce into a cluster. As a result of such processes 
the low-energy products mainly populate the phase space far from midrapidity. 
As another consequence of these secondary interactions we have 
found that the transverse momentum distributions of the produced particles 
decrease versus $P_{T}$ more rapidly around $|y| \approx 4$ than at 
$|y| \approx 0$. 

Actually, the intensive interactions recall the thermalization process, 
therefore, under some conditions thermal models and phenomenologies 
may be applied to describe few characteristics of these reactions. 
In this respect, one can understand 
our results by assuming that the 'kinetic temperature' of baryons at 
midrapidity is much higher than this 'temperature' far from it. 
Therefore, the region outside midrapidity does contribute most strongly 
to the cluster production.

\begin{figure}[tbh]
\includegraphics[width=0.6\textwidth]{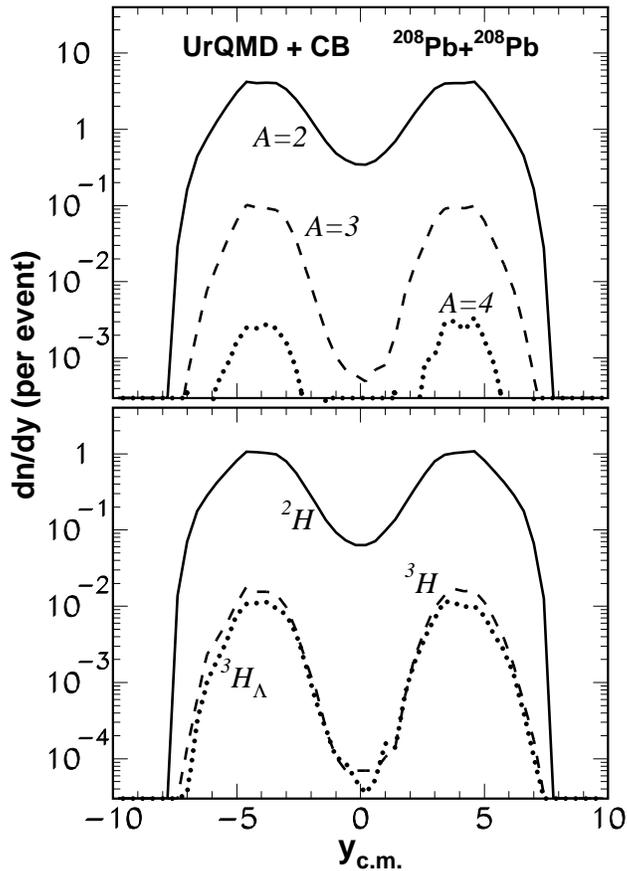}
\caption{\small{ The same as in Fig.~3, 
however, only for normal particles. 
Top panel: light clusters with sum baryon numbers A=2, 3, and 4 (solid, 
dashed and thick dotted lines respectively).
Bottom panel presents rapidity distributions of individual particles: 
deuterons, tritons and hyper-tritons by solid, dashed and thick dotted 
lines respectively.
}}
\label{fig4}
\end{figure}

A more detailed picture of the light fragment production is given in 
Fig.~4. The top panel demonstrates the rapidity distributions of 
normal particle yields with mass (i.e., baryon) numbers of $A=2$, $A=3$, 
and $A=4$. In this case all possible combinations of baryons (including 
both nucleons and hyperons) are taken into account, in order to understand 
the coalescence influence generally. One can see that the yield suppression 
of big fragments is much lager at midrapidity than in the region of the 
maximum fragment yield (at $|y| \approx 4$). For this reason the exploration 
of heavy clusters is more promising at rapidities shifted 
from the midrapidity. This conclusion looks unexpected since more energy 
is deployed in central collisions at midrapidity. The reason is in the 
coalescence mechanism: The constituents should be not only produced, 
they should also have sufficiently low relative velocities to be bound into 
a cluster. 

In the bottom panel of Fig.~4 we show the yields of selected particle 
clusters which 
can be easy identified in the experiment: deuterons ($^2$H), tritons ($^3$H), 
and hyper-tritons ($^3_{\Lambda}$H), versus the rapidity. The distributions 
resemble the same structure as was discussed previously. One can clearly 
see from the figure that the yield ratio of $^2$H to $^3$H is around 
800 at the midrapidity. Note that all calculations on this figure 
are performed for the coalescence parameters $v_c = 0.1 c$, which slightly 
overestimates the deuteron production. Actually, this production and the 
corresponding ratio will be decreased 
by a factor of 2 when we take the more realistic $v_c = 0.07 c$, as is 
clear from Fig.~1. However, one can see that even in the analysed case 
the deuteron-to-triton ratio is decreased to around 60 at $|y| \approx 4$. 
It is also naturally that the yields of $^3$H and $^3_{\Lambda}$H are 
very close, since at such high energy elementary hadron interactions new 
nucleons and hyperons are produced with similar probability. 

The analysis tells us that the region in between the projectile/target 
rapidity and the center-of-mass rapidity is most favorable for the production 
of complex clusters consisting of new produced baryons. We believe that 
experiments should take into account this phase space structure in searching 
for novel exotic nuclear species (including anti-nuclei).
In relativistic heavy ion collisions 
besides the recently observed $^3_{\Lambda}$H nuclei \cite{star,alice-3LH} 
other exotics (like $\Lambda N$, $\Lambda NN$) were under intensive 
discussions \cite{Don15,hyphi-lnn}. 
The extension of measurements into a new rapidity region will increase 
the yields of clusters in the data substantially. It was shown in the 
LHCb experiments 
\cite{LHCb} that not only the midrapidity region but also particles with the 
rapidities around $|y| \approx 4$ can be detected with the 
special detector set-up even at ultra-relativistic energies.

\section{Conclusion}

It was demonstrated that the coalescence process is very important for the 
production of 
light baryonic clusters in ultrarelativistic nuclear collisions. 
We have shown that it is possible to describe spectra of the 
composite clusters 
measured by ALICE at LHC within our UrQMD~+~CB 
approach. We emphasized that the scaling of the elliptic flow of these 
particles may indicate the dominance of the coalescence mechanism. 

The extension of the coalescence results beyond the central collisions 
demonstrate that the maximum yields of such clusters are not located 
at midrapidity. They are essentially shifted toward the 
target and projectile rapidities. This effect reflects the importance of the 
secondary interaction processes which lead to a considerable 
baryon production with low relative momenta. It may also be 
correlated with emerging the hadron fragmentation area. 
Such a new production phenomenon is especially important for forming 
large clusters. Yields of such clusters can be increased by many orders 
while going to the forward/backward region in comparison with the 
midrapidity zone. Here the formation of relatively big exotic, 
hyper- and anti-nuclei becomes very prominent and it is promising for future 
research, as it could provide a unique possibility to study novel nuclear 
species.

\begin{acknowledgments}
A.S.~Botvina acknowledges the support of BMBF (Germany). M.~Bleicher thanks 
the COST Action THOR for support. The authors thank B.~D\"onigus for 
stimulating discussions. 
\end{acknowledgments}

\end{document}